\documentclass[12pt]{article}

\usepackage{scicite}

\usepackage{times}

\usepackage{psfig}

\newenvironment{sciabstract}{%
\begin{quote} \bf}
{\end{quote}}

\newcounter{lastnote}

\title{A Fossil Record of Galaxy Encounters}

\author
{David Elbaz$^{1\ast}$, Catherine J. Cesarsky$^{2}$\\
\\
\normalsize{$^{1}$Service d'Astrophysique/CEA, CE SACLAY, F-91191 Gif-sur-Yvette, 
FR}\\
\normalsize{$^{2}$European Southern Observatory, Karl-Schwarzschild-Strasse 2, 85748 Garching bei Muenchen, Germany}\\
\\
\normalsize{$^\ast$E-mail: delbaz@cea.fr}
}

\date{}

\begin{document} 


\baselineskip24pt


\maketitle 


\begin{sciabstract}
The cosmic infrared background (CIRB) is a record of a large fraction
of the emission of light by stars and galaxies over time. The bulk of
this emission has been resolved by the Infrared Space Observatory
camera. The dominant contributors are bright starburst galaxies with
redshift $z\sim$ 0.8; that is, in the same redshift range as the active
galactic nuclei responsible for the bulk of x-ray background. At the
longest wavelengths, sources of redshift $z\geq$ 2 tend to dominate
the CIRB. It appears that the majority of present-day stars have been
formed in dusty starbursts triggered by galaxy-galaxy interactions and
the build-up of large-scale structures.
\end{sciabstract}


At this very moment, we are receiving light from stars born throughout
 the lifetime of the Universe. Much of this light is in the form of a
 diffuse background about 5\,$\%$ as bright as the cosmic microwave
 background (CMB), a signature of the Big Bang. The Cosmic Background
 Explorer (COBE) satellite--which measured the residual temperature of
 the Big Bang as well as the first fluctuations of density when the
 Universe was only 300,000 years old, the famous seeds of galaxy
 formation--also permitted the first detection of a diffuse background
 due to incipient galaxies emitting light at wavelengths of 100 to
 1000\,$\mu$m, the cosmic infrared background (CIRB). The way galaxies
 evolve from these seeds to present-day galaxies like our own remains
 a mystery, and we do not yet know with certainty the details of the
 cosmic bookkeeping, the global evolution of the total energy emitted
 by stars and galaxies.

The CIRB is a record of a large fraction of the emission of light by
 stars and galaxies over cosmic history. If galaxies formed through
 hierarchical merging, as predicted by current models, then distant
 galaxies may only represent the small precursors of mature galaxies
 like the Milky Way and galaxies in its neighborhood, and galaxy
 formation is a continuous process. Hence, the question "How did
 galaxies form?" may be restated as "When did most of the stars form
 in galaxies?" And another question arises: "Is there any connection
 between the development of large-scale structures and star formation
 within galaxies?" We will see how the information brought about by
 the CIRB and by the studies attempting to determine its origin sheds
 some new light on these questions.
 
\section*{The CIRB}

The x-ray background discovered in 1962 by Giacconi and his
 collaborators during a pioneering rocket experiment was first
 partially resolved into individual sources in the soft energy range
 by the Roentgen X-ray Satellite (ROSAT) (1), then more deeply and in
 a wider energy range by the present-day x-ray observatories Chandra
 and X-ray Multi-Mirror (XMM-Newton) (2-4). Most of the sources are
 active galactic nuclei (AGNs), supermassive black holes in the center
 of galaxies that are accreting matter at a high rate. Recent
 spectroscopic studies of these sources with the Very Large Telescope
 (VLT) at the European Southern Observatory revealed that they mostly
 lie at redshifts (z) below 1, with a mean value of z $\sim$ 0.7 (4). In
 the same way, the light emitted by stars, integrated over time, is
 expected to generate an almost uniform background. In the optical, a
 lower limit to this background was established by calculating the
 integrated contribution of galaxies in the deepest field observed,
 the Hubble Deep Field North (HDFN) (5, 6). The existence of an
 infrared (IR) background in the 5- to 15-$\mu$m wavelength range was also
 predicted (7) but was attributed to the redshifted ultraviolet (UV)
 or optical light from very early galaxies.

In 1983, the first all-sky survey at mid-infrared (MIR) and
 far-infrared (FIR) wavelengths (12 to 100 $\mu$m), performed by the
 Infrared Astronomical Satellite (IRAS), brought about a revolution in
 our understanding of IR emission from local galaxies. Since the IRAS
 data were acquired, we know that in the nearby Universe galaxies
 globally radiate about two-thirds of their light below $\lambda$ = 5
 $\mu$m (i.e., through direct stellar light); the remainder is
 absorbed by dust in the interstellar medium and re-emitted at dust
 temperatures (i.e., in the IR above 5 $\mu$m). Moreover, a new class
 of galaxies was discovered [(8) and references therein] that radiate
 the bulk of their luminosity in the FIR, between 5 and 1000
 $\mu$m. These galaxies, with bolometric luminosities larger than
 10$^{11}$ or 10$^{12}$ solar luminosities, are classified as luminous
 or ultraluminous infrared galaxies (LIRGs or ULIRGs),
 respectively. They produce only 2\,$\%$ of the bolometric luminosity
 density in the local Universe, and the starbursts in them are nearly
 always triggered by galaxy-galaxy interactions. These galaxies must
 have been more numerous in the past, when the Universe was denser and
 galaxies richer in gas. Unfortunately, IRAS was not sensitive enough
 to detect distant objects, but counts of sources at 60 $\mu$m in a
 few deeper fields already showed hints of evolution--that is, an
 increase in the source density or luminosity in the past.

 The extraction of a CIRB from COBE data [(9, 10) and references
 therein], 34 years after the discovery of the x-ray background, was
 almost simultaneous with the introduction of new IR and submillimeter
 observing facilities on the ground [the James Clerk Maxwell Telescope
 (JCMT) and the Institut de Radioastronomie Millimetrique (IRAM) 30-m
 telescope] and in space (the Infrared Space Observatory). The CIRB is
 a measure of the stellar light radiated in the optical and UV (over
 the history of the Universe) that was absorbed by dust and thermally
 reradiated in the IR in the 5- to 1000-$\mu$m range. The energy
 density of this background, about 200 times that of the x-ray
 background and equal to or greater than that of the optical
 background, came as a surprise. It implies that in the past a larger
 fraction of starlight was absorbed by dust and that giant starbursts
 were more common than now. But when? Or at what distance from us?
 When trying to assess at which epoch the Universe was most active in
 the IR, the first clue is the shape of the CIRB spectrum. It is
 reminiscent of the spectral energy distribution of galaxies, as
 observed by IRAS, exhibiting a hump at a wavelength that for
 starburst galaxies or for LIRGs is located at $\sim$ 80 $\mu$m.  With
 a peak intensity of the CIRB around $\lambda$ $\sim$ 140 $\mu$m
 (Fig. 1), and if we assume that the spectral energy density of
 distant starbursts is similar to that of the local ones, then the
 sources responsible for the bulk of the CIRB should be located around
 a redshift of z $\sim$ 0.8; that is, we see them as they were about 7
 billion years ago, when the Universe was about half as old as it is
 today (11). A contribution of more distant galaxies at larger
 wavelengths is suggested by the slope of the CIRB between 300 and
 1000 $\mu$m, which is flatter than the spectral energy distribution
 of a single galaxy at z $\sim$ 0.8 (12). Further studies will be
 needed to identify the sources of the CIRB and to see whether these
 conjectures are confirmed.

\section*{Identification of the Galaxies Responsible for the CIRB}
Ideally, one would wish to observe the IR sky with sufficient spatial
resolution at $\sim$140 $\mu$m to pinpoint the individual galaxies producing
the peak intensity of the CIRB. Unfortunately, this has not yet been
possible. The ISOPHOT detector on board the Infrared Space Observatory
(13) did find a population of galaxies emitting at 170 $\mu$m, which are
one order of magnitude more numerous than expected if the number
density and luminosity of IR galaxies had remained constant with time
(14). The combined contribution of these galaxies to the CIRB amounts
to only $\sim$10\,$\%$ of its value as measured by COBE (14, 15) (Fig. 1),
although fluctuation analysis indicates that fainter sources
contributing to a greater extent to the CIRB are also present in the
ISOPHOT (16, 17) and IRAS images (18). Identifications are difficult
because of the relatively large error box [full width at half maximum
(FWHM) of the point spread function (PSF) = 50 arc sec], but it
appears that the sources detected are either nearby or rare, extremely
bright distant objects.

In the MIR, the gain of sensitivity of the Infrared Space Observatory
with respect to IRAS was more than three orders of magnitude. Deep
surveys at 15 $\mu$m with the camera ISOCAM, also on board the Infrared
Space Observatory, yielded an excess of detections of up to a factor
of 10 with respect to what would be expected if the relevant galaxy
populations had not evolved in the last 10 billion years (19). This
constitutes another proof that the IR luminosity of distant galaxies
and/or their density were much larger in the past than they are
today. Integrating over the ISOCAM source counts, a lower limit to the
CIRB at 15 $\mu$m was established (11) (Fig. 1).  

ISOCAM spectra of local galaxies of all types [(20) and references
therein] show a set of features in the MIR (Fig. 1) that are
attributed to large molecules, probably polycyclic aromatic
hydrocarbons (PAHs) (21) transiently heated to a few hundred
kelvin. These features facilitate the detection by ISOCAM of starburst
galaxies up to redshifts $<$ 1.3 (Fig. 2). These galaxies invariably
have easily identifiable optical counterparts whose IR colors are
indistinguishable from those of optically selected galaxies, but they
exhibit strong H emission (22-25). Their redshift distribution peaks
around z $\sim$ 0.7 to 0.8 (11, 26, 27), as expected if they are
responsible for the bulk of the intensity of the CIRB at its
peak. Their FIR emission was evaluated using the MIR-FIR relation
observed for local galaxies (11, 28). The FIR luminosity of galaxies
correlates strongly with the radio continuum (29), as it does with the
MIR at least up to z $\sim$ 1 (11). It is generally assumed that massive
stars are responsible for the UV photons that heat the IR-emitting
dust and, when they explode as supernovae, for the acceleration of
electrons producing the radio continuum. In the future, the Herschel
satellite will detect these galaxies directly in the FIR up to z $\sim$ 3
(Fig. 2), provided that the spectral energy densities in these distant
galaxies with low metallicity and possibly different distributions of
grain sizes and abundances of polycyclic aromatic hydrocarbons (30,
31) are not too different from the local ones.

 MIR surveys with ISOCAM reach a sensitivity of $\sim$0.1 mJy at 15
$\mu$m; that is, they are able to detect any galaxy producing more
than 20 solar masses of stars per year up to a redshift of z = 1,
hence over the last 60\,$\%$ of the history of the Universe. Using the
MIR-FIR correlations, it is possible to derive a total IR luminosity
for each of the galaxies. Integrating the emissions, it was found that
the galaxies detected in ISOCAM deep and ultradeep surveys are
responsible for about two-thirds of the peak and integrated intensity
of the CIRB. About 75\,$\%$ of these galaxies are LIRGs
($\sim$55\,$\%$) and ULIRGs ($\sim$20\,$\%$) (11); they produce stars
with a median rate of about 50 solar masses per year. As a
consequence, the density of IR luminosity (per unit of comoving
volume) produced by the IR-bright galaxies at z $\sim$ 1 was 70 $\pm$
35 times their present-day luminosity density. This shows that even
though LIRGs and ULIRGs play a negligible role in the local Universe,
they were important actors in the past and represent a common phase in
the evolution of galaxies in general.

An excess of faint galaxies was also detected with the bolometer array
SCUBA on the James Clerk Maxwell Telescope down to the confusion limit
(2 mJy) (32), accounting for about 20\,$\%$ of the CIRB at 850 $\mu$m
[(33) and references therein]. Deeper surveys using gravitational
lensing resolved 60\,$\%$ of the CIRB at 850 $\mu$m into individual
galaxies (33). Unfortunately, the large beam size and the large
redshifts favored by this wavelength range have limited the
identification of the optical counterparts of the bulk of the sources,
and thus the determination of their redshifts, except in rare cases
using interferometry (34-36). In a recent study of bright SCUBA
galaxies with radio counterparts (37) that allow secure
identifications, it was inferred that some of these are indeed
powerful ULIRGs located around z $\sim$ 2. However, the contribution
of sources brighter than 8 mJy to the CIRB is not dominant (38).

Models have been constructed that fit ISOCAM, ISOPHOT, and SCUBA
galaxy counts as well as the CIRB itself (12, 26, 39, 40). There is a
degeneracy in the parameters assumed, defining the relative roles
played by the evolution of galaxies in luminosity and density with
time, but all the models share some general conclusions: About 80\,$\%$ of
the peak of the CIRB at 140 $\mu$m is due to galaxies closer than z = 1.5;
this explains why ISOCAM deep surveys were so efficient in finding the
sources of the CIRB. In contrast, about 70\,$\%$ of the intensity of the
CIRB at 850 $\mu$m is due to galaxies more distant than z = 1.5 (28), of
which SCUBA is already detecting the brightest members. This also
explains why ISOCAM and SCUBA preferentially detect different
populations of galaxies but nonetheless obtain perfectly consistent
results. Overall, 85\,$\%$ of the integrated light of the CIRB can be
attributed to IR luminous galaxies (LIRGs and ULIRGs).

\section*{The CIRB and Large-Scale Structure Formation}

In the local Universe, nearly all ULIRGs are produced by the merging
of two spiral galaxies that will probably result in one
intermediate-mass elliptical galaxy (41, 42). About 75\,$\%$ (43) of the
local ULIRGs already present a luminosity profile following a r1/4
law, typical of early-type galaxies (ellipticals or S0s). The origin
of the starburst phase in LIRGs is less evident, but a recent study of
local objects (44) shows that it is also linked to galaxy environment
ranging from advanced mergers to pairs of spiral galaxies.

In the same vein, less than half of the ISOCAM galaxies exhibit the
disturbed morphology typical of merging galaxies, but it is likely
that tidal interactions or previous encounters triggered the
starbursts, even in the apparently undisturbed ones. The fact that the
integrated contribution of bright starbursts to the cosmic star
formation history or to the CIRB dominates over that of galaxies
forming stars at moderate rates not only implies that most galaxies
must have experienced such a phase in their lifetimes but also
suggests that each of them went through several such phases (39). In
summary, the CIRB appears to be a fossil record of numerous encounters
and/or mergers of galaxies, responsible for their briefly prominent IR
brightness.

An intriguing corollary is that luminous IR galaxies at redshifts
lower than z $\sim$ 1.3 may also be responsible for the formation of the
majority of present-day stars, as well as of heavy elements, in the
local Universe. Indeed, because LIRGs and ULIRGs dominate the cosmic
star formation rate history over that estimated on the sole basis of
direct UV light (26, 28), they should also dominate in the production
of the low-mass stars present today, unless the initial mass function
of stars in these starbursts is strongly depleted of low-mass
stars. Assuming an updated version of the classical Salpeter initial
mass function departing from it below one solar mass (45), the models
of Chary and Elbaz (28)--which fit the CIRB and account for ISOCAM and
SCUBA results--predict that 60\,$\%$ of present-day stars were born nearer
than z $\sim$ 1.3, that is, during the most recent 65\,$\%$ of the age of the
Universe (40\,$\%$ below z $\sim$ 1, 80\,$\%$ below z $\sim$ 2). Because of the dilution
of light in an expanding Universe, it is the galaxies at z $\sim$ 0.8 that
have emitted the bulk of the present-day CIRB. Overall, 80\,$\%$ of the
stars born at z $\leq$ 2 originated in dusty starbursts (LIRGs and
ULIRGs). If these were triggered by galaxy-galaxy interactions, then
the environment of galaxies played a major role in the formation of
present-day stars, as predicted in hierarchical scenarios of galaxy
formation.  

About 68\,$\%$ of the field galaxies from a magnitude-limited sample,
located in a field 8 arc min wide centered on the HDFN, are located in
redshift peaks, whereas all but three of the ISOCAM galaxies in this
field (i.e., 94\,$\%$) belong to these redshift peaks (46, 47), which trace
large-scale structures such as sheets, filaments, and groups or
clusters of galaxies. A structure located at z $\sim$ 0.848 alone contains
almost 30\,$\%$ of the ISOCAM galaxies in the field and includes two AGNs
detected in the x-rays (Fig. 3). At this redshift, the 6-arc min
extension of the structure corresponds to 3 Mpc proper (i.e., too
small to discriminate between a galaxy cluster and a sheet). This
hints at a connection between the formation of large-scale structures
and of galaxies. This also indicates that large-scale structures may
play an important role in the switching on of star formation within
galaxies, but additional MIR deep fields with complete spectroscopic
redshift surveys are obviously required to test the robustness of this
result.

\section*{What Powers the CIRB: Nucleosynthesis or Accretion onto a Black Hole?}

The observed CIRB may originate from light due to nucleosynthesis at
the center of stars or active nuclei (i.e., accretion onto a black
hole). However, detailed studies of the hard x-ray emission of the
ISOCAM galaxies using the deepest x-ray surveys performed with
XMM-Newton in the Lockman Hole and the Chandra X-ray Observatory in
the HDFN have shown that $<$ 20\,$\%$ of their luminosity at 15 $\mu$m is
due to an active nucleus (48). This result is consistent with the
fraction of AGNs within LIRGs and ULIRGs in the local Universe (49,
50). Similarly, the AGNs responsible for the bulk of the x-ray
background were found to produce less than 7\,$\%$ of the
submillimeter background (51). Nonetheless, the redshift and spatial
distribution of ISOCAM galaxies present some striking similarities to
x-ray AGNs. Contrary to optically selected AGNs and x-ray
quasi-stellar objects, the redshift distribution of the Seyfert-type
galaxies responsible for the bulk of the x-ray background also peaks
around z $\sim$ 0.7 (4). Moreover, x-ray AGNs also exhibit strong
clustering, as can be seen in the two deepest images of Chandra, the
Chandra Deep Field South at z = 0.66 and 0.73 (4) and the Chandra Deep
Field North at z $\sim$ 0.843 and 1.017 (52). The structure at z
$\sim$ 0.843 is the same as that mentioned earlier at z $\sim$ 0.848
(Fig. 3). Among the 10 x-ray AGNs detected by Chandra, only two are
also ISOCAM sources.

This suggests that x-ray AGNs and IR luminous galaxies can act as
beacons indicating the regions of growth of large-scale structures. A
similar effect was suggested (53) for the more distant population of
SCUBA galaxies, although this may instead be an artifact of
gravitational lensing (54). The fact that strong starbursts and AGNs
exhibit similar spatial distributions suggests that they represent
successive phases in the life of galaxies. A recent Chandra discovery
(55) may shed new light on this issue: NGC 6240 is a symbiosis between
a typical dusty starburst and an x-ray AGN. Recent Chandra
observations have revealed that this object encompasses in its center
two supermassive black holes probably in the process of merging. NGC
6240 may therefore represent the missing link between dusty starbursts
and x-ray AGNs.
 
\section*{Conclusions and prospects}
The recent extraction of a CIRB from the data obtained by the COBE
satellite, combined with the results of deep surveys in the IR and
submillimeter range, has revealed the importance of star formation in
strong starbursts in the history of the Universe. The cosmic star
formation rate density was more than one order of magnitude larger
about 7 billion years ago (z = 0.8) than it is today (28). More than
75\,$\%$ of this evolution is due to dusty starbursts (LIRGs and ULIRGs)
that produced stars at a mean rate of $\sim$50 solar masses per year at the
earlier epoch. Although the peak and the bulk of the CIRB can be
attributed to galaxies at relatively modest redshifts (z $\leq$ 1.3), more
distant galaxies dominate the emission at submillimeter wavelengths,
to which their intrinsic emission is redshifted because of the
expansion of the Universe. The brightest of these galaxies, ULIRGs at
redshifts z $\geq$ 2, are being detected at 850 $\mu$m from ground by bolometer
arrays at the focus of radio telescopes. The overall importance of
ULIRGs seems to have been even greater in those earlier times.

The rapid star formation revealed by IR observations may be connected
to large-scale structures. There is a similarity in the redshift
distributions, and possibly also the clustering properties, of the
bright starburst galaxies and x-ray-selected AGNs. The existence of a
link between the triggering of a starburst phase in galaxies and the
fueling of a central black hole, already suggested by the study of
local ULIRGs discovered by IRAS (56), is supported by this independent
evidence.

These findings can also be summarized by noting that galaxies,
paradoxically, are sociable and shy at the same time. They are
sociable because they brighten up in company. They are shy because
during their encounters with other objects, the UV light of their
newly formed stars is absorbed by dust and thermally re-emitted in the
IR, so that they blush.

The fecundity of this topic promises a bright future for the next
generation of IR instruments such as the Space Infrared Telescope
Facility (SIRTF), which will be able to bridge the gap between ISOCAM
and SCUBA and to study LIRGs and ULIRGs in the 1 $\leq$ z $\leq$ 2 redshift
range. Later, the PACS instrument on the Herschel telescope will
resolve the CIRB directly in the FIR, and the James Webb Space
Telescope with its MIR camera MIRI will permit detailed studies of the
individual sources. The fluctuations of confusion-limited surveys with
Herschel will also provide the opportunity to obtain information on
FIR sources at redshifts so high that they cannot be detected
individually (40). Finally, the combination of all these instruments
with the high spatial resolution images and spectra of the Atacama
Large Millimeter Array (ALMA) is likely to bring about a new
revolution in our understanding of how stars and galaxies form.





\begin{thebibliography}{2}
\bibitem{notescuba}
SCUBA, with a beam of 14.7 arcsec FWHM, is limited by diffraction to a
confusion limit of about $\sim$ 2 mJy. The shape of the spectra of
ULIRGs favors the detection of very distant objects, with $z\geq$
2. This is why, paradoxically, SCUBA misses most of ISOCAM sources,
but can detect ULIRGs at $z\geq$ 2 invisible for ISOCAM (Fig. 2).
\bibitem{notebkg}
$\sim$80\,$\%$ of the background light is due to galaxies fainter
than 2 mJy at 850\,$\mu$m.
\end{thebibliography}


\clearpage

\noindent

\begin{figure}
\psfig{file=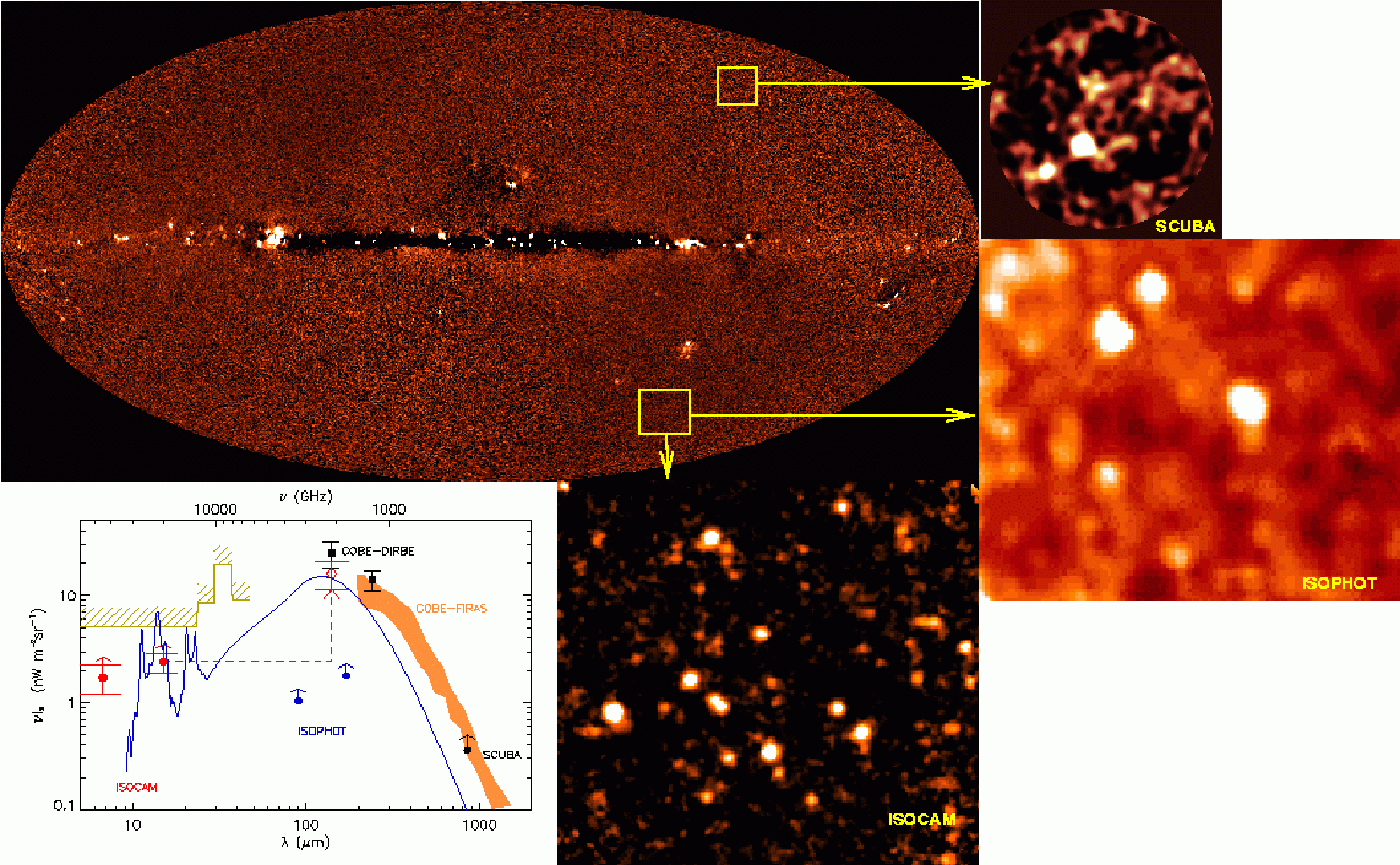,width=16cm,height=12cm}
\caption{The cosmic IR background. (Upper left) Map of
the full sky as seen in IR light at wavelength 140 $\mu$m from the
instrument DIRBE (Diffuse Infrared Background Experiment) on board
COBE. (Upper right) Deep image (down to 2 mJy) of the HDFN with SCUBA
at 850 $\mu$m [(57), resolution = 12 arc sec]. (Lower right) 15- and
170-$\mu$m images of a region of 9 arc min in the southern sky (Marano
FIRBACK field) from the ISOCAM (58) and ISOPHOT (14) instruments on
board the Infrared Space Observatory (resolutions of 4.6 and 52 arc
sec, respectively). (Lower left) Intensity of the CIRB as a function
of wavelength and frequency. The solid squares with error bars and the
orange area give the actual intensity of the CIRB from the DIRBE and
FIRAS instruments on board COBE, respectively. The dots with upward
arrows (see references in the text) are lower limits set by galaxy
counts from ISOCAM (6.75 and 15 $\mu$m), ISOPHOT (90 and 170 $\mu$m),
and SCUBA (850 $\mu$m). The lower limit set by ISOCAM at 15 $\mu$m was
used to compute a lower limit to the CIRB at its peak around 140
$\mu$m (dashed arrow) using the MIR-FIR relation (11). The spectral
energy density is that of a typical LIRG normalized to the 15-$\mu$m
point and redshifted to z = 0.8. It exhibits broad features attributed
to polycyclic aromatic hydrocarbons and peaks at about 80 $\mu$m (in
the rest frame). The hatched area is an upper limit set by TeV -ray
photons that annihilate with MIR photons through electron-positron
pair production (59-61).}
\end{figure}

\begin{figure}
\psfig{file=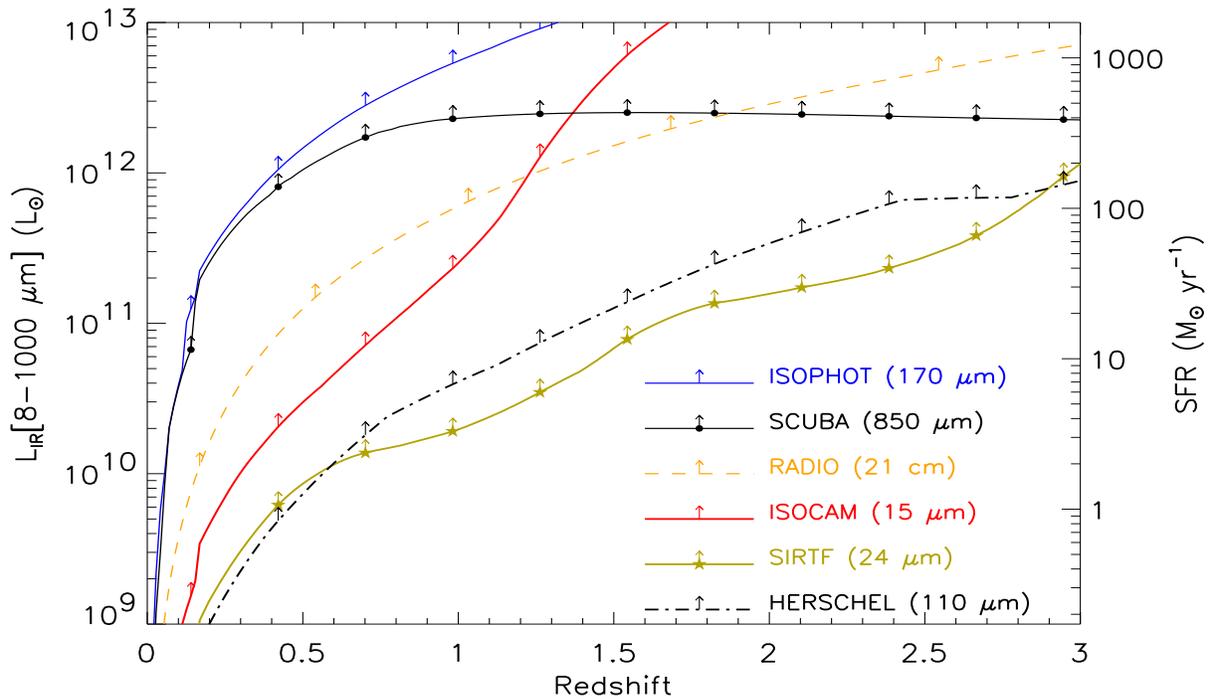,width=16cm,height=10cm}
\caption{IR luminosity (left axis) and star formation rate (SFR,
right axis) as a function of redshift corresponding to the 5-$\sigma$
sensitivity (S) limits at different wavelengths ($\lambda$) from
ISOCAM ($\lambda$ = 15 $\mu$m, S = 0.1 mJy) and the VLA in the radio
(62) ($\lambda$ = 21 cm, S = 40 $\mu$Jy), and to the confusion limits
of ISOPHOT ($\lambda$ = 170 $\mu$m, S = 120 mJy), of SCUBA ($\lambda$
= 850 $\mu$m, S = 2 mJy), and of the future spatial experiment MIPS on
board SIRTF (63) ($\lambda$ = 24 $\mu$m, S = 22 $\mu$Jy, GOODS Legacy
Program) and HERSCHEL-PACS ($\lambda$ = 110 $\mu$m, S = 5.1 mJy).}
\end{figure}

\begin{figure}
\psfig{file=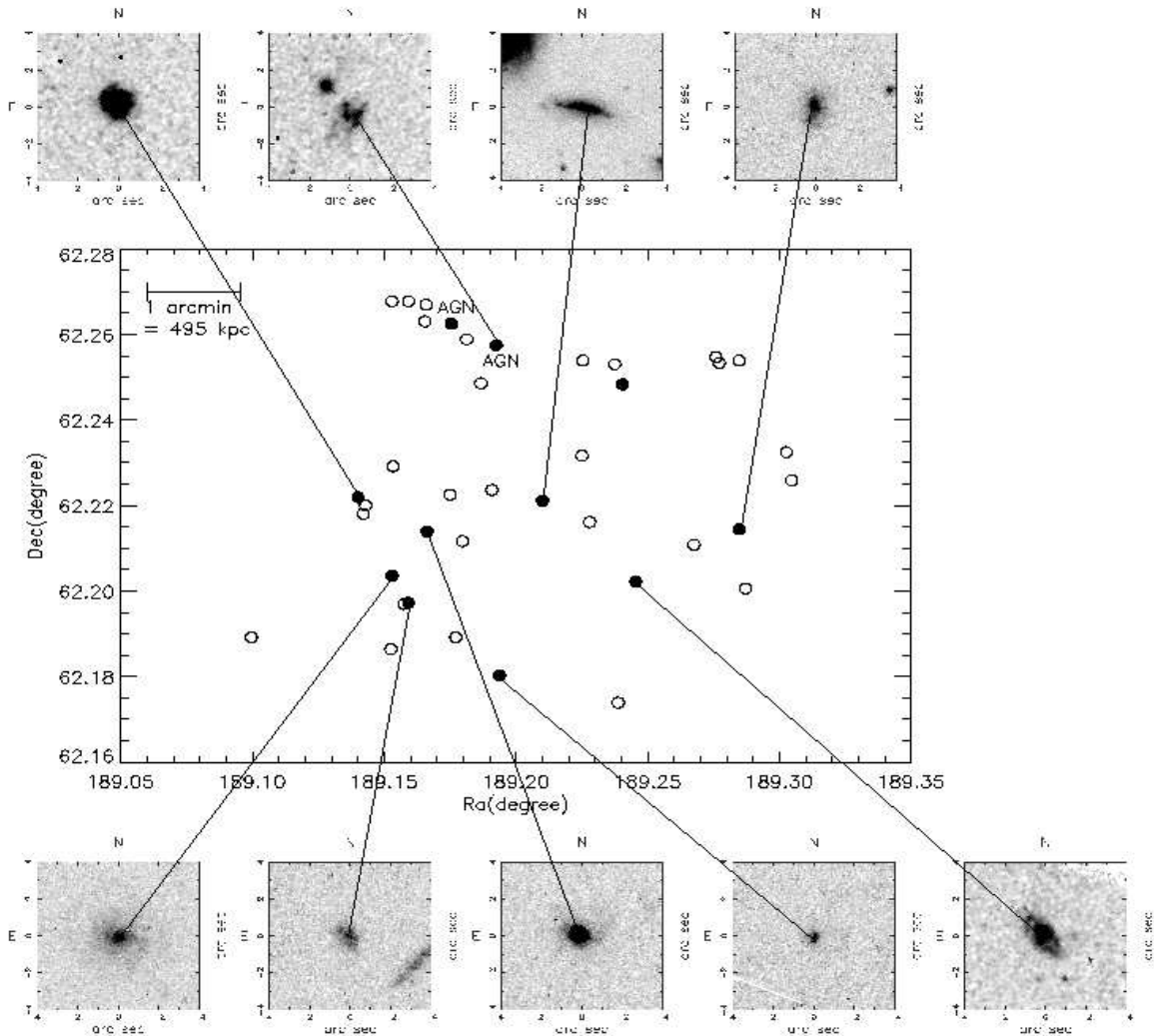,width=16cm,height=14cm}
\caption{A large-scale structure at z = 0.848 (3 Mpc proper
diameter). Empty circles are field galaxies; dark circles are
15-$\mu$m ISOCAM galaxies. Postage-stamp HST images of the ISOCAM
galaxies are shown when available (from the DEEP archive
database). The positions of two active nuclei (AGNs) are
indicated. This is the highest concentration of dusty starbursts ever
detected. Each ISOCAM galaxy is forming stars at a rate of about 50
solar masses per year.}
\end{figure}

\end{document}